\documentstyle[12pt, psfig]{article}
\renewcommand{\thepage}{\arabic{page}}
\newcommand{\nc}{\newcommand}
\nc{\beq}{\begin{equation}} \nc{\eeq}{\end{equation}}
\nc{\beqa}{\begin{eqnarray}} \nc{\eeqa}{\end{eqnarray}}
\nc{\lsim}{\begin{array}{c}\,\sim\vspace{-21pt}\\< \end{array}}
\nc{\gsim}{\begin{array}{c}\sim\vspace{-21pt}\\> \end{array}}

\newcommand{\drawsquare}[2]{\hbox{%
\rule{#2pt}{#1pt}\hskip-#2pt%  left vertical
\rule{#1pt}{#2pt}\hskip-#1pt%  lower horizontal
\rule[#1pt]{#1pt}{#2pt}}\rule[#1pt]{#2pt}{#2pt}\hskip-#2pt%  upper horizontal
\rule{#2pt}{#1pt}}% right vertical

\newcommand{\Yfund}{\raisebox{-.5pt}{\drawsquare{6.5}{0.4}}}%  fund
\newcommand{\Yasymm}{\raisebox{-3.5pt}{\drawsquare{6.5}{0.4}}\hskip-6.9pt%
        \raisebox{3pt}{\drawsquare{6.5}{0.4}}}%  antisymmetric second rank

\newcounter{mysection}
\newcounter{mysubsection}
\newcommand{\mysection}[1]{\stepcounter{mysection}\setcounter{equation}{0}
\setcounter{mysubsection}{0}\par\bigskip\noindent{\large\bf
\themysection .\ #1}\nopagebreak[4]\par\vskip .3cm}

\newcommand{\mysubsection}[1]{\stepcounter{mysubsection}
\par\medskip\noindent{\large\it
\themysection .\themysubsection\ #1}\nopagebreak[4]\par\vskip .3cm}
\def\l:{\mathopen{:}\,}
\def\r:{\,\mathclose{:}}

\def\inbar{\,\vrule height1.5ex width.4pt depth0pt}
\font\cmss=cmss12 \font\cmsss=cmss12 at 7pt
\def\IZ{\relax\ifmmode\mathchoice
{\hbox{\cmss Z\kern-.4em Z}}{\hbox{\cmss Z\kern-.4em Z}}
{\lower.9pt\hbox{\cmsss Z\kern-.4em Z}}
{\lower1.2pt\hbox{\cmsss Z\kern-.4em Z}}\else{\cmss Z\kern-.4em
Z}\fi}
\def\IB{\relax{\rm I\kern-.18em B}}
\def\IC{{\relax\hbox{$\inbar\kern-.3em{\rm C}$}}}
\def\ID{\relax{\rm I\kern-.18em D}}
\def\IE{\relax{\rm I\kern-.18em E}}
\def\IF{\relax{\rm I\kern-.18em F}}
\def\IG{\relax\hbox{$\inbar\kern-.3em{\rm G}$}}
\def\IP{\relax{\rm I\kern-.18em P}}
\begin{document}

\begin{titlepage}

{\hbox to\hsize{hep-th/9810010 \hfill UCSD-PTH-98-34}}
{\hbox to\hsize{September 1998 \hfill }}
\bigskip

\begin{center}

\vspace{.5cm}

\bigskip

\bigskip

\bigskip

{\Large \bf   On the  One Loop Fayet-Iliopoulos Term in 

\vspace{.3cm}

Chiral Four Dimensional Type I Orbifolds }

\bigskip

\bigskip

\bigskip

{\bf Erich Poppitz}
\smallskip

{\tt  epoppitz@ucsd.edu}

\bigskip

\bigskip

{\small \it Department of Physics \\
University of California at San Diego\\
9500 Gilman Drive\\
La Jolla, CA 92093, USA}\footnote{After January 1, 1999: 
Department of Physics, Yale University, New Haven, 
CT 06520-8120, USA.}

\bigskip
 
\bigskip

{\bf Abstract}
\end{center}
 
We consider the generation of Fayet-Iliopoulos
terms at one string loop in some recently found $N=1$ open string orbifolds 
with anomalous $U(1)$ factors with nonvanishing trace of the charge.  
Low-energy field theory arguments lead one to expect a one loop
quadratically divergent Fayet-Iliopoulos term.
We show that a one loop  
Fayet-Iliopoulos term is not generated, due to a cancellation 
between contributions of worldsheets of different topology. 
The vanishing of the one loop Fayet-Iliopoulos term in open string 
compactifications is related to the cancellation of twisted 
Ramond-Ramond  tadpoles.
\end{titlepage}

\renewcommand{\thepage}{\arabic{page}}

\setcounter{page}{1}

\mysection{Introduction}

The choice of ground state of string theory is one of the greatest
obstacles to its phenomenological application. The breaking of spacetime 
supersymmetry and  stabilization  of the moduli remain two of the most 
pressing problems in this regard \cite{GSW}. 
Most of the studies of supersymmetry breaking in the framework of string
theory in the past have been in closed (heterotic) string
compactifications.
In recent years, however,  it has been realized that the 
previously known five distinct string theories are related by various dualities.
In particular, open string (type I) theories were found to be dual to
closed string (heterotic) theories \cite{S}.
 In view of the newly found dualities, it is natural to
ask whether the weakly coupled description of the 
world might be in terms of one of the other, dual string theories. It is,
therefore, of interest to investigate compactifications of type I theories 
and study their properties as well.

The first four dimensional type I orbifold with $N=1$ supersymmetry
and chiral matter content was found in \cite{ABPS}. Since then, following
\cite{PS}, \cite{GP}, a  number of other examples have been 
constructed  \cite{exs}. The occurrence of anomalous $U(1)$ factors
of the gauge group is common in these examples. The anomaly is cancelled,
as in the closed string case, by a generalized Green-Schwarz mechanism, but
there are important differences \cite{DM}, \cite{BLPSSW}, \cite{IRU}; 
see the discussion at the end of the Introduction and in the concluding Section.

It is well known that
$N=1$ supersymmetry in four dimensions allows for 
Fayet-Iliopoulos (FI) terms\footnote{To avoid confusion, 
in all our formulae,  we
include the factor of $1/g^2$ ($g$ is the gauge coupling; recall that 
$g^2 \sim g_{string}$ for open strings)  in the gauge superfield kinetic term; 
hereafter, 
by ``FI term" we always mean the value of $\zeta_{FI}^2$ in the lagrangian
(\ref{FI}), with this normalization of the vector superfield.} 
 for the $U(1)$ factors of the gauge group
 in the lagrangian, of the form:
\beq
\label{FI}
{\cal{L}}_{FI} ~=~ \zeta_{FI}^2 ~ \int d^4 \theta ~V~.
\eeq 
FI terms can also play an important 
role in supersymmetry breaking \cite{WB}, \cite{N}.
Moreover, in the case of anomalous $U(1)$ factors with 
nonvanishing trace of the charge,  ${\rm Tr}~ Q_{U(1)} \ne 0$, 
a FI term can be generated at one loop (a 
nonrenormalization theorem forbids the appearance of 
higher loop corrections \cite{FNPRS}, \cite{DSW}, \cite{W}).
The one loop contribution to the FI term is quadratically divergent,
$\zeta_{FI}^2 \sim \Lambda_{UV}^2 {\rm Tr}~ Q_{U(1)}$, and is therefore
highly cutoff and regularization dependent; for example, it vanishes in
dimensional regularization.
It does, therefore, represent a ``matching"
 contribution, and should be calculated in the underlying theory, 
valid beyond the scale of the ultraviolet cutoff, $\Lambda_{UV}$.

Anomalous $U(1)$ factors with Tr $Q_{U(1)} \ne 0$ also appear in
closed string compactifications \cite{DSW}. It is a well established fact,
that at one closed string loop level a FI term is always generated
\cite{DIS}, \cite{ADS}, \cite{L}, \cite{DL}.
The torus amplitude leads to a finite contribution to the FI term, with
the ultraviolet cutoff replaced by the inverse string scale,
$\Lambda_{UV} \sim 1/\sqrt{\alpha^\prime}$.

Here we will investigate the generation of one loop FI terms in 
open string compactifications. An indication that the open string
case might  differ from the closed one is provided by 
considering the way  the quadratic divergence is regulated 
in  string theory graphs. The one loop FI term in field theory is
proportional to $\int d^4 p/p^2$, and represents a quadratically divergent
contribution. In string theory, it is replaced by 
\beq
\label{oneloopquadratic}
\int ~{d^4 p \over p^2} \rightarrow \int ~d^4 p ~\int_{t_{min}}^\infty~d t ~
e^{- t p^2}~,
\eeq
where $t$ denotes the  modulus of the relevant worldsheet. 

In the
closed string case, $t \sim {\rm Im}~ \tau$, where $\tau$ is the  
modular parameter
of the torus. Since each torus is counted only once (due to modular
invariance), the integration
over $\tau$ is restricted to the fundamental region 
$|\tau| \ge1$, ${\rm Im} ~\tau > 0$, 
$|{\rm Re}~ \tau| \le 1/2$; hence the lower
limit of integration in (\ref{oneloopquadratic}) never vanishes, $t_{min} > 0$.
This implies that momenta of order the inverse string scale and higher 
are always cut off from the momentum integral 
(all scales in (\ref{oneloopquadratic})  are in
terms of the string scale, $\alpha^\prime$, which we set, 
hereafter, equal to $1/2$).

In the case of the open string, $t$ also represents the 
modulus of the relevant worldsheet---the
cylinder or M\" obius strip. Since there is
no modular invariance, 
the lower limit of the modulus integral in (\ref{oneloopquadratic})   is zero,
 $t_{min} = 0$. Therefore the high loop-momentum contributions
are not cut off and the divergent 
parts must cancel between different graphs (since we believe that string
theory is finite). 

In the rest of the paper, we will show that this cancellation does, in fact, occur.
We perform the one loop open string calculation of the 
 FI term in a type I compactification
in the orbifold limit. We will concentrate on the simplest case, the $T^6/Z_3$ 
orbifold
of type I theory \cite{ABPS}. 
We will show that the contributions of the cylinder and M\" obius
strip cancel exactly. 
Therefore, there is no FI term generated at the one loop 
level. We will see that the vanishing of the one loop FI term is closely  
related\footnote{This not surprising---it is well-known \cite{GSW} 
that UV divergences in the one-loop 
open-string channel have a dual interpretation as
 IR divergences in the tree-level closed-string channel.}
 to the cancellation of  twisted Ramond-Ramond tadpoles, 
which is important for consistency of the 
orbifold \cite{GP}.

Our main conclusion is that the only contribution to the FI term in
the open string compactifications considered here occurs at tree-level
and is due to giving expectation values to the orbifold blow-up modes
(these contributions have been discussed first in \cite{DM} and recently
in \cite{IRU}). Since the FI term now is due to the expectation
value of a modulus, its value is, at least in perturbation theory, arbitrary, rather
than the fixed value (of order the string scale) obtained in the heterotic case.
In the final section of the paper, we 
consider these tree-level contributions in 
some more detail and point out some 
implications for supersymmetry breaking.
We believe, based on the heuristic arguments given 
above, that our result is more generally valid; we leave its extension to other
type I compactifications for future work.

\mysection{The calculation}

\mysubsection{The $Z_3$ orbifold of type I theory}

In this section, we will describe in some detail the orbifold, introduce
our notation, and 
find explicitly the Chan-Paton factors of the massless open string
modes that survive the orbifold projection. These will be useful in
the calculation  of the subsequent sections
and in the interpretation of the results.

The $Z_3$ orbifold of type I theory was first constructed in ref.~\cite{ABPS}. 
We will describe the construction in the more familiar language of ref.~\cite{GP}.
To describe the orbifold, and further in our calculation, we 
use light-cone variables $X^\pm = (X^0 \pm X^9)/\sqrt{2}$. The compact
directions (the six-torus $T^6$) 
are $X^1, \ldots, X^6$. 
To construct the $Z_3$ orbifold of type I theory,  one mods out the Hilbert
space of type I theory by the action of a $Z_3$ discrete symmetry.
We take 
 the operator 
\beq
\label{alpha}
\hat{\alpha} ~=~ \exp \left\{- ~{4  \pi  i \over 3}
 \left( \hat{J}_{12} ~+~ \hat{J}_{34} ~+~ \hat{J}_{56} \right) \right \}~
\eeq
to generate
the $Z_3$ symmetry. Here
$\hat{J}_{ij}$ are the generators of rotations in the $ij$-plane ($i,j = 1,\ldots ,8$).
Note that, with this definition, $\hat{\alpha}^3 = 1$ in the spinor representations 
as well.\footnote{Note that our definition of the 
orbifold $\hat{\alpha}$ (\ref{alpha}) 
 is equivalent to the often used
$\hat{\alpha}^\prime = 
\exp \left[- (2 \pi i/3) \left( J_{12} + J_{34} - 2 J_{56}\right)\right]$, which also
obeys $(\hat{\alpha}^\prime)^3 = 1$  in the spinor representation.} 
We will  employ the unhatted letter $\alpha$ to denote 
$e^{2 \pi i/3}$. The transverse (to the light-cone) noncompact
 directions are 
$X^7$ and $X^8$. The orbifold action (\ref{alpha}) leaves 
invariant four supercharges
($N=1$ in four dimensions).

The construction of the closed string
sector of the orbifold proceeds by 
projecting out closed string states that are not invariant 
under the action of $\hat{\alpha}$; in addition, one has to add twisted sectors. 
This results in the appearance of 27 chiral multiplets in the twisted sector
of the orbifold (one chiral multiplet for each of the $3^3 = 27$ fixed points)
\cite{ABPS}. 
The construction of the open string sector of the orbifold 
proceeds along the lines of \cite{DM}, \cite{GP}. One finds that 
the orbifold group also acts on the Chan-Paton factors 
$\lambda$, as 
\beq
\label{cpaction}
\hat{\alpha}: ~ \lambda \rightarrow \gamma_\alpha \lambda \gamma_\alpha^{-1}~.
\eeq 
Here $\gamma_\alpha$ is an embedding of the orbifold group in the 
gauge group, whose explicit form we will give below. $\gamma_\alpha$
 has to satisfy the tadpole cancellation conditions \cite{GP}, which, in this case,
amount to  requiring Tr $\gamma_\alpha = - 4$. 

The massless open string states that survive the orbifold projection can
be found by considering the action of the orbifold group on the massless vertex 
operators. 
The massless boson open string vertex operator in type I
theory is (we use light-cone gauge and Green-Schwarz fermions; our notations
are identical to those of \cite{GSW}):
\beq
\label{vxop}
V( \xi, k; w) ~= ~\xi^i ~\left( \dot{X}^i - R^{ij} (w) ~k^j \right) ~ e^{i k \cdot X(w)}~
 {\hat{\lambda}}~,
\eeq
where $R^{ij}(w) = (1/4) S^a (w) \gamma^{ij}_{ab} S^b(w)$, and $S^a$ are
the Green-Schwarz fermions transfoming in the ${\bf 8_c}$  spinor 
representation of $SO(8)$ \cite{GSW}.
Here $\hat{\lambda}$ represents the relevant Chan-Paton factor.
Gauge bosons in the four dimensional theory have polarization vectors,
whose transvere components are along
the $X^7$, $X^8$ directions, while 
the polarization vectors for scalars are along
the $X^1 \ldots X^6$ directions. We will take as an example 
\beq
\label{polarization}
\xi_1 ~= ~{1 \over \sqrt{2} }~(0,~ 1, ~-i, ~0, \ldots , 0)~, ~~ \xi_2 ~=~
 ~{1 \over \sqrt{2} }~(0,~1, ~i, ~0, \ldots, 0)~,
\eeq
 where $\xi_1$ and $\xi_2$
correspond to a complex scalar ($\xi_2$) 
and its complex conjugate ($\xi_1$); the components of the polarization 
vectors are listed in order $X^0, X^1, ..., X^9$. 

The massless open string spectrum is now found by requiring that the 
massless state 
vertex operators be invariant under the combined action of $\hat{\alpha}$
(\ref{alpha}) on
the fields in (\ref{vxop}) and on the Chan-Paton factor $\hat{\lambda}$
(\ref{cpaction}). Using the tadpole cancellation condition to determine
the form of $\gamma_\alpha$, one finds that
the
 Chan-Paton factors of the states invariant under the orbifold projection
 are as 
described explicitly below.

We use a basis where the
$SO(32)$ Chan-Paton factors
are real antisymmetric matrices.
The action  of the orbifold group on the Chan-Paton factors (see
eq.~(\ref{cpaction})), is 
represented by the $SO(32)$ matrix:
\beqa
\label{gamma}
\gamma_\alpha ~=~ \left( \begin{array}{ccc}
{\bf 1}_{8} & 0  & 0 \\
0& -{1 \over 2} ~ {\bf 1}_{12}& {\sqrt{3} \over 2} ~ {\bf 1}_{12}\\
0&  - {\sqrt{3} \over 2} ~ {\bf 1}_{12}& -{1 \over 2} ~ {\bf 1}_{12}
  \end{array} \right) ~,
\eeqa
obeying the tadpole consistency condition Tr $\gamma_\alpha = - 4$.
Here we use a block-matrix notation: ${\bf 1}_k$ denotes a ($k\times k$)
dimensional unit matrix. 

To find the gauge fields' Chan-Paton
factors, we note that vertex operators with polarization vectors along the
noncompact directions are invariant under the action of $\hat{\alpha}$. Hence
the gauge Chan-Paton factors should obey 
$\lambda = \gamma_\alpha \lambda \gamma_\alpha^{-1}$. Thus, we
find
\beqa
\label{cpgauge}
\lambda_{gauge}~=~ \left( \begin{array}{ccc}
{\bf A}^\prime &  0  & 0 \\
0 & {\bf A}  & {\bf S} \\
0 & - {\bf S} & {\bf A }\end{array} \right) ~,
\eeqa
where $ {\bf A}^\prime$ is an ($8 \times 8$) antisymmetric 
matrix that generates $SO(8)$, while ${\bf A}$ and ${\bf S}$ are ($12 \times 12$)
antisymmetric and symmetric matrices, respectively. ${\bf A}$ and ${\bf S}$ 
 together 
generate $U(12)$, so that ${\bf A} - i {\bf S}$, ${\bf A} + i {\bf S}$ are
 the antihermitean 
generators of the fundamental and antifundamental representations,
respectively. For further use, note that the 
anomalous $U(1)$ factor is generated by the trace part of the symmetric tensor.
From eq.~(\ref{cpgauge}) it follows that the gauge group of the four dimensional
theory is $SO(8)\times U(12)$.

The Chan-Paton factors for the matter fields are
complex antisymmetric matrices (recall that the matter fields are complex
linear combinations of gauge bosons polarized in the compact directions; see
eqs.~(\ref{vxop}, \ref{polarization})). 
Since under the action of $\hat{\alpha}$, the field dependent
part of the scalar vertex operator (denoted by $V^\prime$)
 with polarization vector $\xi_1$, see eq.~(\ref{polarization}),
 transforms as $\hat{\alpha} V^\prime \hat{\alpha}^{-1} =
\alpha^{-1} V^\prime$, in order for the state to be invariant,
 its Chan-Paton factor $\lambda^\dagger$ has to
satisfy $\gamma_\alpha \lambda^\dagger
\gamma_\alpha^{-1} = \alpha \lambda^\dagger$, 
and is explicitly given by:
\beqa
\label{cpmatter}
\lambda_{matter}^\dagger~=~ \left( \begin{array}{ccc}
0 & {\bf  b}  & i~{\bf b} \\
- {\bf b}^T & {\bf a} & - i~{\bf a} \\
- i~ {\bf b}^T & - i ~{\bf a} & - {\bf a} \end{array} \right) ~. 
\eeqa
Here we use the same block structure as in (\ref{cpgauge}): 
${\bf b}$ is an ($8 \times
12$) dimensional matrix, representing the fields that transform under both
$SO(8)$ and $U(12)$, 
while ${\bf a}$ is an ($12 \times 12$) antisymmetric tensor.
The complex conjugate matter fields are represented by the 
hermitean conjugate
matrix $\lambda_{matter}$.  
It is easy to check that under gauge transformations, 
$\delta \lambda_{matter}^\dagger = \left[ \lambda_{gauge},
\lambda_{matter}^\dagger \right]$, the factors ${\bf b}$ and ${\bf a}$ transform 
(in terms of finite transformations of $SO(8)$, ${\bf O} = e^{{\bf A}^\prime}$,
 and $U(12)$,  ${\bf U} = e^{{\bf A} - i {\bf  S}}$) as:
${\bf a} \rightarrow {\bf U a } {\bf U}^T$,
${\bf b} \rightarrow {\bf O b} {\bf U}^\dagger$. 
Thus, the massless matter content of the orbifold in terms of chiral
superfields is given by three copies of $(\Yfund, \overline{\Yfund}) +
({\bf 1}, \Yasymm)$ under $SO(8) \times U(12)$, corresponding
to the three complex compact directions. The $U(1)$ charges of
the fields ${\bf a}$ and ${\bf b}$ are $2$ and $-1$, respectively.

It is easily seen, then, that the $U(1)$ is anomalous, 
and that Tr $Q_{U(1)} \ne 0$.
Thus, low-energy field theory considerations \cite{N, FNPRS, DSW, W} 
lead one to expect a one loop quadratically divergent FI term. 
The following sections are devoted to the string calculation of this term.

\mysubsection{Factorization of the four-point amplitude}

In this section, we will describe the idea behind the calculation
of the one loop FI term in string theory. The main points below 
are as in
 the closed string calculation; we will be correspondingly brief and
only mention the essential differences.

Adding a FI term (\ref{FI}) to the lagrangian leads to scalar 
masses---restoring the canonical normalization of
the vector superfield kinetic term and expanding the 
anomalous $U(1)$ D-term potential, see \cite{WB},
one finds mass terms for all scalar fields, 
proportional to their $U(1)$ charges: $m^2_\phi \sim g^2 q_{\phi} \zeta_{FI}^2$. 
It is these mass terms  
that  are easiest to compute in string theory \cite{DSW}, 
\cite{DIS}, \cite{ADS}, \cite{DL}. 

\begin{figure}[ht]
\vspace*{13pt}
\centerline{\psfig{file=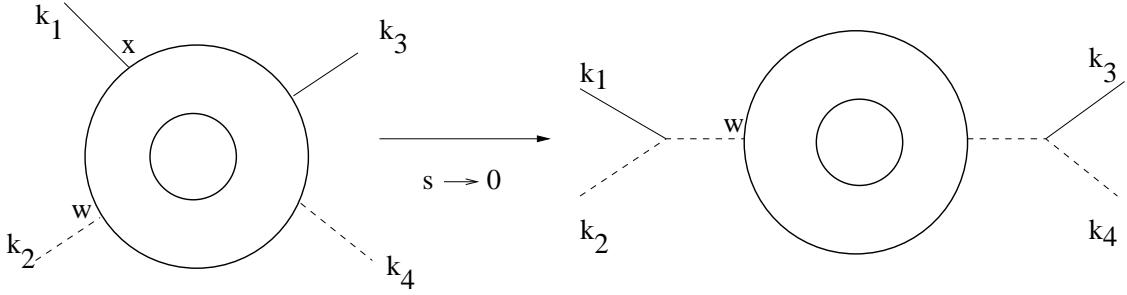}}
\vspace*{13pt}
\caption{\small {Factorization of the four-point amplitude of gauge 
boson$-$scalar (scalars are denoted by a dashed line) 
scattering in the $s$-channel ($s = -(k_1 + k_2)^2  = -2 k_1 \cdot k_2$);
$x$ and $w$ denote the points the vertex operators are attached to the
world sheet (\ref{opescalarvector}). A summation over the four 
noncyclic permutations, which allow for $s$-channel poles, accounts for the 
correct Chan-Paton factor of the state exchanged in the $s$-channel.}}
\end{figure}

Since string amplitudes are on-shell quantities, 
it is natural to define masses as 
poles in scattering amplitudes at appropriate values of the external momenta.
To compute the one loop contribution to the 
scalar masses in our orbifold, consider the elastic 
scattering amplitude of a $U(1)$ gauge boson and a scalar (see Fig.1).
Let $k_1$ and $k_2$ be the momenta of  the incoming gauge boson and scalar,
respectively, and $k_3$ and $k_4$---of the outgoing. One expects
a pole in the amplitude at $s = m^2$, where $m$ is the mass of any 
scalar that can be produced in the $s$-channel.
In other words, one expects 
$A \sim 1/(s  - m^2) \simeq m^2/s^2$; the second equality is true
if $m$ is treated as a perturbation around zero tree-level
mass (i.e. as a  ``mass insertion").
To see this factorization explicitly, consider 
the OPE of a scalar and gauge boson vertex operators:
\beq
\label{opescalarvector}
V_{gauge} (\xi_g, k_1; x) ~ V_{scalar} (\xi_m, k_2; w) ~=~ 
(\xi_g \cdot k_2) ~ \left( {w \over x -w} \right)^{1 - k_1 \cdot k_2}~ 
: V_{scalar} (\xi_m, k_1 + k_2; w) : ~+~ \ldots~.
\eeq
The following identity is useful in deriving 
eq.~(\ref{opescalarvector}), as well as eq.~(\ref{scalarope}) from Sect. 2.3:
{\flushleft{$
{1\over 2}~ ( a_1 \cdot \gamma_{ab} \cdot b_1 )~ ( a_2 \cdot \gamma_{bc} 
\cdot b_2 ) - {1\over 2}~
( a_1 \cdot \gamma_{cb} \cdot b_1 )~ ( a_2 \cdot \gamma_{ba} 
\cdot b_2 ) ~=~  $}} 
\beq\label{identity}
- (a_1 \cdot b_2) ~ (a_2 \cdot \gamma_{ac} \cdot b_1)
+ (b_1 \cdot b_2) ~ (a_2 \cdot \gamma_{ac} \cdot a_1)
+ (a_1 \cdot a_2 )~ (b_2 \cdot \gamma_{ac} \cdot b_1)
- (b_1 \cdot a_2) ~ (b_2 \cdot \gamma_{ac} \cdot a_1)~.
\eeq
The matrices $\gamma_{ab}^{ij}$ are as defined in \cite{GSW}.
In (\ref{opescalarvector})
 $\xi_g$ ($\xi_m$) are the polarization vectors of the gauge 
boson  (scalar)  and the dots denote terms that 
do not contribute to the factorized amplitude
in the $s \rightarrow 0$ limit.
Inserting (\ref{opescalarvector}) into the  four-point amplitude
\cite{GSW}, and integrating over $x$, 
using 
\beq
\label{integral}
{\rm lim}_{k_1 \cdot k_2 \rightarrow 0} ~\int_w^1 {d x \over x} 
~\left({w \over x -w} \right)^{1 - k_1 \cdot k_2}~ F(x) ~ = ~ {1 \over k_1 \cdot
k_2}~ F(w) ~+ ~{\rm nonsingular}~,
\eeq
one finds that
the four-point  amplitude in the limit $s \sim k_1 \cdot k_2 \rightarrow 0$ 
factorizes (as shown on Fig.~1; see also \cite{DIS}, \cite{ADS}) into the
product of a 
tree-level amplitude for producing a scalar (the $\xi_g \cdot k_2$ factor
in (\ref{opescalarvector})),  a massless scalar propagator (the $1/k_1 \cdot k_2$
factor due to (\ref{integral})), a two point scalar amplitude, 
with the scalar vertex operators taken off-shell 
(as on the r.h.s. of eq.~(\ref{opescalarvector})), another massless scalar
propagator, and another tree-level amplitude for producing a scalar; 
we note that factorization also holds for the surface with a crosscap.

From the above discussion, we conclude that
the two-point scalar amplitude plays the role of
 mass insertion in field theory. The one loop correction
to the mass is therefore found by computing the two-point scalar
amplitude with the scalars taken slightly off-shell, and by taking the limit 
$k^2 \rightarrow 0$ at the end of 
the calculation \cite{DIS}, \cite{ADS}.\footnote{
There is one difference from the closed string case here. When
considering the factorization of the four-point amplitude in the $s$-channel,
we have to sum over all noncyclic permutations of the external lines
which allow for $s$-channel poles (i.e. over four of the six noncyclic 
permutations
of the four-point amplitude) and include a trace over the external Chan-Paton
factors in the  order corresponding to each permutation.
It is easy to see that each permutation generates a minus sign, 
and, consequently, that the  Chan-Paton factor accompanying 
each of the two scalar vertex operators in the ``mass insertion" is equal to 
$[ \lambda_{U(1)}, \lambda_{scalar}]  \cite{GSW}$, in our case---the gauge 
transformed scalar Chan-Paton factors; 
see discussion after eq.~(\ref{cpmatter}).}

\mysubsection{Vanishing of the two-point scalar amplitude}

As argued in the previous section, 
  the scalar mass squared is related to the one loop 
scalar two-point function, computed slightly off-shell, with the on-shell
limit taken in the end of the calculation. 
The computation of the scalar two-point function and the demonstration of
its vanishing is the subject of this section.  The two-point one-loop
scalar amplitude is of the same order, ${\cal{O}}(g_{string}) \sim g^2$, as 
 the field theory one-loop contribution to the scalar mass: the 
annulus and M\" obius strip have $\chi = 0$, while
each vertex operator contributes a factor of $ g_{string}^{1/2} \sim g$.

To calculate the two-point scalar amplitude, we will use the operator formalism
with Green-Schwarz fermions. The expression for the two-point
function in type I theory  has the form \cite{GSW}:

{\flushleft{$ A(k_1, k_2) ~ =$} }
\beq
\label{2ptfn}
~ 
\int_0^1  {d w \over w} ~ \int_w^1 {d x \over x} ~\int {d^{10} p \over (2 \pi)^{10}}  ~
{\rm Tr} \left[ ~ {1 + \hat{\alpha} + \hat{\alpha}^2 \over 3} ~
V_1 (\xi_1, k_1; x) ~ {1 + \hat{\Omega}\over 2} ~ V_2 (\xi_2, k_2; w) ~
{1 + \hat{\Omega}\over 2} ~ w^{\hat{L}_0} ~\right] ~.
\eeq
Here $V(\xi, k; x)$ are the vertex operators for the scalars, see eqs.~(\ref{vxop},
\ref{polarization}); 
they include the Chan-Paton factors appropriate for the 
states under consideration---the matrices of the form 
(\ref{cpmatter}), which we
denote $\lambda_1^\dagger$ and $\lambda_2$ for $V_1$ and $V_2$ 
in (\ref{2ptfn}), respectively.

The trace in eq.~(\ref{2ptfn}) is 
over all states in the Hilbert space of type I string theory. 
The operator $(1 + \hat{\alpha} + \hat{\alpha}^2)/3$ 
performs a projection on $Z_3$ invariant states.
In the formalism we are using  \cite{GSW}, 
the Hilbert space is a tensor product of
three spaces: the space of zero modes, times the Fock space of $X^i$ and $S^a$ 
oscillators, times the space of Chan-Paton indices (all states are also labeled
by the loop momentum, the trace over which is factored out in (\ref{2ptfn})). 
The various operators in (\ref{2ptfn}) act on one, or on several, components
of the tensor product, as discussed below.
 Thus, the trace in (\ref{2ptfn}) reduces to a sum of products of traces in each of
these spaces.
The operator $\hat{L}_0 \equiv {\hat{p}^2 \over 2} + \hat{N}$, where $\hat{N}$ 
is the excitation number operator.  Here $p$ is momentum of the string 
running in the loop. For a compact direction of radius $R$, we have to replace
the continuum momentum integral in (\ref{2ptfn}) 
by a sum over momentum modes: 
$\int d p \rightarrow (1/R) \sum_n$. 
 $\hat{\Omega} = - (-)^{\hat{N}}$ is the 
orientation reversal operator. It is important for what follows that the 
operators $\hat{\alpha}$ and $\hat{\Omega}$ act on the Chan-Paton indices
as well; we will include the overall minus sign in $\hat{\Omega}$ in its action
on the Chan-Paton factors.

Because of the trace over the supersymmetric zero modes
 (denoted hereafter by ${\rm Tr}_0$), 
the terms proportional to $\dot{X}$ in the vertex operators
(\ref{vxop}) do not contribute to trace (using the formulas for
zero-mode traces of \cite{GSW}, and the definition of $\hat{\alpha}$
(\ref{alpha}), it is easy to see that ${\rm Tr}_0 ~ \hat{\alpha} = 0$). 
Hence, only the pieces involving the
Green-Schwarz fermions in the vertex operators (\ref{vxop}) 
contribute to the trace in
(\ref{2ptfn}). 
It is clear, then, that the amplitude will be proportional to
$k_1 \cdot k_2$.  Now, 
because of momentum conservation $k_1 = - k_2 = k$; hence, 
in the limit when the two scalars are on shell, 
$k_1 \cdot k_2 = - k^2 = 0$, the two-point
amplitude would naively vanish. 
However, as in the closed string case, the loop integral 
in (\ref{2ptfn}) can contribute a $1/k^2$ factor, which cancels the overall
$k^2$ factor and leads to a nonvanishing result as the two scalars are
taken on-shell \cite{DIS, ADS, DL}.
To see this, consider the OPE of two scalar vertex operators 
(see (\ref{identity})):

{\flushleft{$V_{scalar} (\xi_1, k_1; x) ~ V_{scalar} (\xi_2, k_2; w) ~ = ~ $}}
\beq
\label{scalarope} ~
{1 \over 4} ~
(k_1 \cdot k_2) ~ \left( {w \over x -w} \right)^{1 - k_1 \cdot k_2}~ 
\xi_1^i ~\gamma_{a b}^{i j} ~\xi_2^j ~ S^a (w) ~ S^b (w) ~
: e^{i (k_1 + k_2) \cdot X(w)} : ~+~ \ldots~,
\eeq
where, as usual, the dots denote terms which do not contribute to the 
$1/k^2$ pole.
Inserting (\ref{scalarope}) into (\ref{2ptfn}) and 
integrating over $x$, eq.~(\ref{integral}), we see that the $1/k^2$ factor
cancels the overall $k^2$ factor in (\ref{scalarope}).\footnote{
We also note that the contribution to the amplitude of the term with 
$\hat{\Omega}$ between the two scalar vertex operators vanishes---the 
OPE of $V_1(x) \hat{\Omega} V_2 (w)$ has a singularity at $x = -w$, 
(this can be seen using $\hat{\Omega} V(w) \hat{\Omega} = V(-w)$); 
however, this is outside the region of integration in (\ref{2ptfn}) and 
does not contribute a $1/k^2$ pole.}
Thus, the calculation of the  nonzero contribution
to the two point function reduces to:
\beq
\label{twopointreduced}
A(k^2 \rightarrow 0) ~ = ~{\rm const.} ~
\int_0^1 {d w \over w} \left( { 2 \pi \over \log w} \right)^2 ~{\rm Tr}~
\left( {1 + \hat{\alpha} + \hat{\alpha}^2 \over 3} ~ R^{ij} (w)~ {1 + \hat{\Omega}
\over 2} ~w^{\hat{N}} \right) ~\xi_1^i ~\xi_2^j ~,
\eeq
where we absorb various inessential factors in the overall constant, as 
mentioned below.
 In obtaining eq.~(\ref{twopointreduced}), we have performed the $x$-integral
as described above; see eq.~(\ref{integral}). We have also
 performed the loop-momentum integral over the four noncompact
momenta, resulting in the factor of $(2 \pi/\log w)^2$. 
It is important to note, that since only the terms with  $\hat\alpha$
and $\hat{\alpha}^{2}$ in (\ref{2ptfn}) are nonvanishing (because of
the zero-mode trace, as explained in the
following paragraph), 
only  vanishing  momentum modes 
 in the compact directions  contribute to the trace (we have 
omitted the overall $1/(2 \pi R)^6$ 
factor resulting from them, as well as the overall
momentum conserving delta function).

Now we go on to the computation of the remaining trace in
(\ref{twopointreduced}). 
We begin with considering the  zero-mode traces.
To this end, first note that
the terms without insertions of $\hat{\alpha}$ do not contribute---the
corresponding traces over the space of zero modes
 vanish, ${\rm Tr}_0 R^{ij}_0 = 0$ (recall that one
needs an insertion of at least 
eight zero modes to get nonvanishing zero-mode 
trace \cite{GSW}). Here 
$R^{ij}_0 = (1/4) S^a_0  \gamma^{ij}_{ab} S^b_0$, where $S_0^a$ denote
the zero modes of the Green-Schwarz fermions \cite{GSW}. Furthermore,
because of unbroken supersymmetry, it is also easy to see (as mentioned
above) that ${\rm Tr}_0 \hat{\alpha} = 0$ ($\hat{\Omega}$ acts
trivially on the zero modes). Therefore  both fermion fields in 
(\ref{twopointreduced}) have to correspond to zero modes. Hence, the
only zero-mode traces we need to compute are:
 \beq
\label{zeromodetrace}
{\rm Tr}_0 ~ \hat{\alpha} ~ R^{12}_0 ~=~ - 3 \sqrt{3}/2 ~ ~{\rm and} ~ ~
{\rm Tr}_0 ~ \hat{\alpha}^2 ~ R^{12}_0 ~=~ 3 \sqrt{3}/2 ~, 
\eeq
where, for brevity, we have only given
 the $R_0^{12}$ component, relevant for our choice of 	
polarization vectors (\ref{polarization}).
The traces are computed using the formulae for
the action of the rotation generators
on the zero-mode states, given in \cite{GSW},
and the definition of $\hat{\alpha}$, eq.~(\ref{alpha}); we also note
that, in view of our result, we only make use of the relative minus
sign in  (\ref{zeromodetrace}), which can be obtained by
computing the trace of the transposed matrix
and using $(R^{ij}_0)^T = - R^{ij}_0$, and $\alpha^T = \alpha^2$.

\begin{figure}[ht]
\vspace*{13pt}
\centerline{\psfig{file=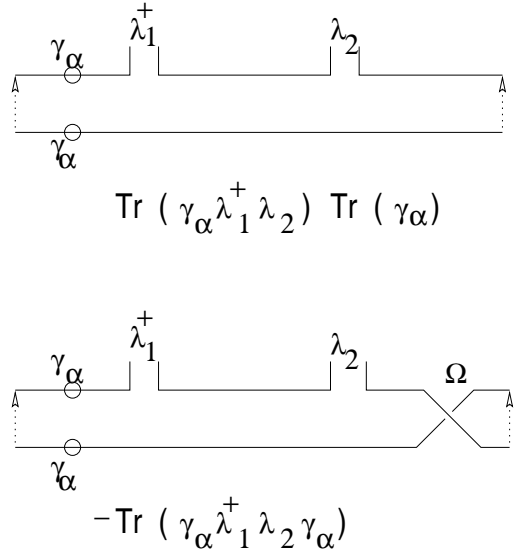}}
\vspace*{13pt}
\caption{\small{
A schematic representation of the two worldsheets 
(the ends should be glued along the arrows), contributing to
the $1/k^2$ pole in the scalar two point function. Insertions 
of $\gamma_\alpha$
are denoted by a circle. The Chan-Paton traces 
Tr~($\hat{\alpha} \hat{\lambda}_1^\dagger 
\hat{\lambda}_2$) and Tr~($\hat{\alpha} \hat{\lambda}_1^\dagger 
\hat{\lambda}_2 \hat{\Omega}$) are computed by tracing the product of
the matrices $\gamma_\alpha$ and $\lambda_{1,2}$ in the order they appear 
along each boundary of the worldsheet, and are shown under each 
worldsheet. A similar representation follows for
the trace with $\hat{\alpha} \rightarrow \hat{\alpha}^{2}$ and amounts to
replacing $\gamma_\alpha \rightarrow \gamma_\alpha^{-1}$. As explained in 
the text, 
because of the unbroken space-time supersymmetry, 
the term without insertions of $\hat{\alpha}$ in the trace does not contribute. } }
\end{figure}

Now we go on to the traces over the Chan-Paton indices. 
We define the action of $\hat{\alpha}$ (and, similarly, of  $\hat\alpha^{2}$) 
on the Chan-Paton indices, 
${\hat{\alpha}} \vert a b \rangle =   
\vert c d \rangle (\gamma_{\alpha})_{c a } (\gamma_{\alpha})_{d b}$, while 
the Chan-Paton factors from the vertex operators
$\hat\lambda$ of (\ref{vxop}) act on, say,
the first indices only: 
${\hat{\lambda}} \vert a b \rangle =  \vert c b\rangle 
\lambda_{c a}$, since
they are both attached to the same boundary; note that
 the  above definition preserves
the correct cyclic order, and is consistent with the action of $\hat\alpha$ on
the Chan-Paton factors (\ref{cpaction}).
For the evaluation of 
the M\" obius strip, we also need the action of 
$\hat{\Omega}$ on the Chan-Paton indices: $\hat{\Omega} \vert
a b\rangle = - \vert b  a \rangle$, where, as mentioned above, 
we include the overall minus
sign in the action on the Chan-Paton indices.
Finally, $\langle c d \vert a b 
\rangle = \delta_{c a} \delta_{d b}$. Thus, 
the Chan-Paton trace of the unit operator
gives a factor of $n^2$ ($n$ being the number of $D$-branes), while
the trace of $(1 + \hat{\Omega})/2$ gives $(n^2 - n)/2$ (with the above 
definition of $\hat\Omega$ appropriate for an $SO(n)$ projection; for us, $n=32$). 
In computing the various Chan-Paton traces, where appropriate, 
we also make use of
$\gamma_\alpha^T = \gamma_\alpha^{-1}$, $\gamma_\alpha^2 =
\gamma_\alpha^{-1}$. The computation of the relevant---the
ones with insertions of $\hat\alpha$ and $\hat{\alpha}^2$---Chan-Paton traces
yields:
\beqa
\label{cptraces}
{\rm Tr}_{C.-P.} \left( \hat{\alpha} ~\hat{\lambda}_1^\dagger ~
\hat{\lambda}_2 \right)~&=& ~
{\rm Tr} \left( 
\gamma_\alpha ~\lambda_1^\dagger ~\lambda_2
 \right)~ {\rm Tr} \left(
 \gamma_\alpha \right) ~\nonumber \\
{\rm Tr}_{C.-P.} \left( \hat{\alpha} ~\hat{\lambda}_1^\dagger~ \hat{\lambda}_2
~\hat{\Omega} \right)~&=& ~- ~
{\rm Tr} \left( 
\gamma_\alpha ~\lambda_1^\dagger~ \lambda_2 ~\gamma_\alpha \right)~ .
\eeqa
Eq.~(\ref{cptraces}) can be verified using the rules outlined above (which
are also graphically represented on Fig.~2). Similar relations hold for the traces
where $\hat\alpha$ is replaced by $\hat{\alpha}^2$---one simply replaces
$\gamma_\alpha$ by its inverse in (\ref{cptraces}).

Finally, we consider the trace over all nonzero modes---the Fock space of 
$X^i$ and $S^a$
oscillators. It is easy see that these two traces precisely cancel, because of
the unbroken space-time supersymmetry. For example, the trace over the 
$X^i$ oscillators, Tr $\hat{\alpha} ~w^{\hat{N}}$, yields a factor of  
$\prod_{n=1}^{\infty} 
(1 - \alpha ~w^n)^3 (1 - \alpha^{-1} ~w^n)^3 (1 - w^n)^2$ in the 
denominator,
while the trace over the fermionic nonzero
modes yields the same factor in the numerator; similarly, the contribution
of the $X^i$ oscillators to Tr $\hat{\alpha}~
 \hat{\Omega} ~w^{\hat{N}}$
 is precisely cancelled by the contribution of the $S^a$ oscillators. 
Thus, we conclude that
no massive string excitations contribute to the two-point function, 
as in the closed string case, and in accord with field theory expectations.

Now we can assemble the contributions 
(\ref{zeromodetrace}, \ref{cptraces})
to the two-point amplitude (\ref{twopointreduced}).  We write the
amplitude (\ref{twopointreduced}) as $A = A_{C} + A_{MS}$, omitting
the overall constant.
Note that the contributions of the Chan-Paton
and zero-mode traces factorize and can be combined (below we use
eq.~(\ref{polarization}) for $\xi_1$ and $\xi_2$). The cylinder contribution
$A_{C}$ (the term 
without $\hat\Omega$ in (\ref{twopointreduced})) to 
the two point amplitude is then, using (\ref{zeromodetrace}, \ref{cptraces}):
\beqa
\label{cyl} 
A_{C} ~&= &~ 
 {\rm Tr} \left[ \left( - i \gamma_\alpha + i \gamma_\alpha^{-1} \right)
\lambda^\dagger_1  \lambda_2 \right]~{\rm Tr} \left( \gamma_\alpha \right)~
\int_0^1 {d w \over w} \left( { 2 \pi \over \log w} \right)^2~\nonumber \\
&=&~- ~4  ~
 {\rm Tr} \left[ \left( - i \gamma_\alpha + i \gamma_\alpha^{-1} \right)
\lambda^\dagger_1 \lambda_2 \right]~
\int_0^1 {d w \over w} \left( { 2 \pi \over \log w} \right)^2~,
\eeqa
where we used the tadpole cancellation condition,
Tr $\gamma_\alpha = - 4$; see (\ref{gamma}).
The contribution $A_{MS}$ of  the M\" obius
strip (the term with $\hat\Omega$ in (\ref{twopointreduced}))
to the two-point amplitude can also be evaluated by assembling the
traces discussed above, 
eqs.~(\ref{zeromodetrace}, \ref{cptraces}), with the result:
\beqa
\label{ms}
A_{MS} ~&=&~ {\rm Tr}  \left[ (i \gamma_\alpha^2 - i
\gamma_\alpha^{-2}) \lambda^\dagger_1 \lambda_2 \right] ~
\int_0^1 {d w \over w} \left( { 2 \pi \over \log w} \right)^2 ~ \nonumber \\
&=& ~ {\rm Tr}  
\left[ (-i \gamma_\alpha + i \gamma_\alpha^{-1} )
 \lambda^\dagger_1 \lambda_2  \right] ~ 
\int_0^1 {d w \over w} \left( { 2 \pi \over \log w} \right)^2~,
\eeqa
where we used $\gamma_\alpha^2 = \gamma_\alpha^{-1}$ in the last line.

Now, we observe that both the cylinder (\ref{cyl}) and
M\" obius strip (\ref{ms}) 
contributions are proportional to the same
Chan-Paton traces. Using the explicit representations of
the gauge Chan-Paton factors (\ref{cpgauge})
on the one hand, and  of $\gamma_\alpha$ (\ref{gamma}) on
the other, we see that 
$\lambda_{U(1)} \sim (\gamma_\alpha - \gamma_\alpha^{-1})$; 
recall that $U(1)$ is generated by the trace part of the symmetric
tensor in (\ref{cpgauge}).  
Therefore, the Chan-Paton traces in (\ref{cyl}, \ref{ms}) are
the same ones that appear in the tree-level 
coupling of the $U(1)$ gauge boson to
scalars.  Thus, as expected (see the beginning of Section 2.2),  
both the cylinder and M\" obius strip 
contributions to the scalar masses are proportional to their  $U(1)$ charges.

However, both contributions, eqs.~(\ref{cyl}, \ref{ms}), are badly ultraviolet
divergent---in a way of
comparison with eq.~(\ref{oneloopquadratic}), we note that
the modular parameter in (\ref{cyl}, \ref{ms}) is 
$t \sim \log w$ and the region of unsuppressed
large loop momenta corresponds to $w = 1$. 
Moreover, the divergent integrals in (\ref{cyl}) and (\ref{ms}) 
come with different coefficients, 
and the contributions appear not to cancel. 
There is, however, a subtlety in adding divergent contributions 
from Riemann surfaces of different topology: as discussed in  
\cite{GSW}, one has to rescale the modular parameter of the M\" obius
strip relative to the cylinder before adding the divergent contributions.
The correct rescaling of 
the variable of integration in (\ref{ms}) is $w = (w^\prime)^{1/4}$ \cite{GSW}. 
While there appears to be
 no first-principle derivation of this rescaling, it is precisely the 
relative scaling between
the cylinder and M\" obius strip contributions that is needed in order to
obtain anomaly cancellation in type I theory for $SO(32)$ gauge group
(rather than the inconsistent choice $SO(8)$); as argued in
 \cite{GSW}, the rescaled variables are more natural, since the Green's 
functions on the cylinder and M\" obius strip are simply related, when 
expressed in the rescaled variables, and their 
contributions to scattering amplitudes can be combined into a single 
integral.\footnote{More physically, this rescaling is the one 
consistent with unitarity: the poles in the tree-level 
cross-channel in both graphs then correspond to the masses of the 
closed string excitations. We thank Z. Bern for discussions on 
this.}
This is also the rescaling used to derive the tadpole consistency 
conditions \cite{GP}. It amounts to a particular (consistent in all known cases) 
regularization of the two divergent integrals  in (\ref{cyl}, \ref{ms}).

It is clear then, by inspecting eqs.~(\ref{cyl}) and (\ref{ms}), that the
cylinder and M\" obius strip contributions to the scalar masses---and
hence, to the FI term---precisely cancel upon 
rescaling the modular parameter in (\ref{ms}),
$w = (w^\prime)^{1/4}$: $A = A_{C} + A_{MS} = 0$. 
The factor of $- 4$ in the cylinder contribution (\ref{cyl}), which appeared 
from the tadpole consistency condition, is cancelled by the factor of $4$ coming
from the rescaling in (\ref{ms}).
 We see, therefore,  that the vanishing of the
one loop FI term is closely related to the cancellation of the twisted
Ramond-Ramond tadpoles.

\mysection{Summary and discussion}

Our main result  is that there is no FI term induced at the one loop level
in type I orbifolds with anomalous $U(1)$s with Tr $Q_{U(1)} \ne 0$, contrary
to the low-energy field theory expectation. 
The vanishing of the one-loop
contribution occurs because of a cancellation between the contributions of
worldsheets of different topology, and is closely related to the vanishing
of the twisted Ramond-Ramond tadpoles.\footnote{We 
note that the situation here is different from
compactifications of type I on smooth Calabi-Yau manifolds; see
the recent discussions in \cite{MR}, \cite{IRU}.}

We conclude  with mentioning another important difference
between open and closed string compactifications with anomalous $U(1)$
factors. In the closed string case, it is the model-independent axion
(which is in the same supermultiplet with the dilaton) that transforms
nonlinearly under the anomalous $U(1)$ and appears in the 
Wess-Zumino terms canceling the anomaly \cite{DSW}. In the open-string
compactifications that we consider here (and, generally, on 
D-branes on orbifolds), the
role of axions is played by model-dependent fields: the Ramond-Ramond 
twisted scalars from the closed string sector transform nonlinearly under
the $U(1)$ and participate in the relevant Wess-Zumino terms. This was
first noted in \cite{DM} and was recently discussed in \cite{IRU}. 
The twisted Ramond-Ramond scalars appear in the same supermultiplet as
the blow-up modes of the orbifold. In chiral superfield notations their kinetic
lagrangian is:
\begin{equation}
\label{RRlagr}
\int ~d^4 \theta ~\left( C ~ + ~ C^\dagger ~ + V \right)^2 ~ + ~ \ldots ~,
\end{equation}
where dots denote higher-order terms.
The leading 
term (\ref{RRlagr}) can be written by demanding
$U(1)$ invariance and a smooth  kinetic term for
 $C$ in the orbifold limit $\langle C \rangle = 0$. 
Here $C$ is the superfield whose imaginary part shifts under the anomalous
$U(1)$, while $V$ is the $U(1)$ vector superfield; in addition to (\ref{RRlagr}),
the field $C$ also has a Wess-Zumino coupling to the gauge field strengths, of
the form $\int d^2 \theta C W^\alpha W_\alpha$ \cite{DM} (in (\ref{RRlagr}) 
various constants have been set to one). 
In a superunitary gauge, the term (\ref{RRlagr}) represents a mass term
(of order the string scale)  for the anomalous $U(1)$ vector superfield. By 
giving an expectation value to the real part of $C$ (blowing up the orbifold)
one can induce ``tree-level" FI terms, with 
$\zeta_{FI}^2 \sim  \langle C + C^\dagger \rangle$, as follows from (\ref{RRlagr}).
 
That (\ref{RRlagr}) is correct follows from the computation of  
ref.~\cite{DM} of  the coupling of the real part of $C$ (the twisted NS-NS field)
to the D-term of the vector superfield (and from a subsequent 
supersymmetry transformation). This coupling
arises from the disk with two scalar vertex operators attached to the boundary,
and a closed string twisted NS-NS scalar vertex operator in the bulk \cite{DM},
and is of the same order in the string coupling,  
 ${\cal{O}}(g_{string}) \sim g^2$,  as the would-be one loop 
contributions (\ref{cyl}, \ref{ms}).

The vanishing of the one loop FI term has some implications 
for supersymmetry breaking in the low-energy field theory of the 
$T^6/Z_3$ orbifold, considered recently in \cite{LPT}. As 
was argued there, in the model with 
appropriate discrete Wilson lines (leading to an $U(5) \subset SO(8) \times
U(12)$ theory), 
supersymmetry breaks regardless of the value of the FI term 
(for fixed values of the closed string modes: as usual, supersymmetry
breaking  is plagued by a runaway problem). 
On the other hand, in the $U(4)$  theory, which is continuously connected
to the $SO(8)\times U(12)$ theory considered here, 
supersymmetry breaking depends crucially on the value of the FI term.
Hence, since the one loop FI term vanishes, in the orbifold limit supersymmetry 
in the $U(4)$ theory is unbroken.

We can summarize our result by saying that 
the only contribution to the FI term in open string
orbifolds comes at tree level,
 by giving expectation values to the 
twisted sector NS-NS fields (i.e. by blowing up the orbifold), due to the
coupling (\ref{RRlagr}). A separate term in the lagrangian, 
of the form $\int d^4 \theta V$, allowed by gauge invariance,
and  expected to appear at one loop in the low-energy 
field theory, is not present in the effective action.  

We expect that our result is more generally valid, rather than hold just for the
$T^6/Z_3$ orbifold. In particular, we expect it to be valid
for noncompact constructions involving $D3$ branes at orientifold
fixed planes as well (i.e. on $\IC^3/\Gamma$ orientifolds); for the $\IC_3/Z_3$
 case  this conclusion follows from our calculation by 
$T$-dualizing and taking the  limit of large compactification
radius. It would be interesting to 
extend the result, either by explicit computation,
or, possibly, by some general argument, to other (e.g. including also
five branes) open string compactifications
 with anomalous $U(1)$s.

\bigskip

It is a pleasure to thank S. Trivedi for conversations that lead to this
investigation, and the Aspen Center for Physics for hospitality. 
Discussions with Z. Bern, S. Chaudhuri, K. Intriligator, C. Johnson, 
E. Silverstein, and  K. Skenderis are also gratefully acknowledged.

\nc{\ib}[3]{ {\em ibid. }{\bf #1} (19#2) #3}
\nc{\np}[3]{ {\em Nucl.\ Phys. }{\bf #1} (19#2) #3}
\nc{\pl}[3]{ {\em Phys.\ Lett. }{\bf B#1} (19#2) #3}
\nc{\pr}[3]{ {\em Phys.\ Rev. }{\bf D#1} (19#2) #3}
\nc{\prep}[3]{ {\em Phys.\ Rep. }{\bf #1} (19#2) #3}
\nc{\prl}[3]{ {\em Phys.\ Rev.\ Lett. }{\bf #1} (19#2) #3}

\end{document}